\documentclass[twocolumn,pra,superscriptaddress,nofootinbib]{revtex4-2}
\usepackage{cancel}
\usepackage{ulem}
\usepackage{soul}
\usepackage{graphicx}
\usepackage{dcolumn}
\usepackage{bm}
\usepackage{amsmath} 
\usepackage{hyperref}
\usepackage{xcolor}
\usepackage{orcidlink}
\usepackage{ulem}

\hypersetup{
	colorlinks=true,
	linkcolor=blue,
	citecolor=blue
}
\begin{document}

\preprint{APS/123-QED}

\title{Experimental investigation of Lévy flights for step-length distributions with a length-dependent local power exponent.}

\author{Isaac C. Nunes\,\orcidlink{0009-0000-0513-2289}}
\affiliation{Departamento de Física, CCEN, Universidade Federal da Paraíba, Caixa Postal 5008, 58051-900 João Pessoa, Paraíba, Brazil}
\affiliation{Departamento de Física, Universidade Federal de Pernambuco, 50670-901 Recife, Pernambuco, Brazil}

\author{Jesús P. López\,\orcidlink{0000-0002-3145-8351}}
\affiliation{Departamento de Física, CCEN, Universidade Federal da Paraíba, Caixa Postal 5008, 58051-900 João Pessoa, Paraíba, Brazil}
\author{Thierry Passerat de Silans\,\orcidlink{0000-0003-1765-8354}}
\email{thierry@otica.ufpb.br}
\affiliation{Departamento de Física, CCEN, Universidade Federal da Paraíba, Caixa Postal 5008, 58051-900 João Pessoa, Paraíba, Brazil}

\date{\today}

\begin{abstract}
Light propagating in a resonant atomic vapor performs a Lévy flight random walk, in which rare long steps dominate the transport. In contrast to the ideal power-law step-length distributions usually assumed in the literature, the actual distribution in atomic vapors is characterized by a local Lévy index $\alpha(\ell)$ that varies smoothly with the step length $\ell$. Moreover, the walker alternates between two distinct step-length distributions, depending on whether an atomic collision occurs during each scattering event. In this work, we experimentally investigate which effective Lévy index is retrieved from diffuse transmission measurements in such a nonideal Lévy flight, and how the alternation between the two distributions influences the transmission. From measurements in a dense cesium vapor for different system sizes and atomic densities, we show that the measured Lévy index corresponds to the local exponent evaluated at the system size, $\alpha_T=\alpha(\ell=L)$. Surprisingly, the measured values are consistent with the step-length distribution associated with atomic collisions, that has heavier tails, even when the collision probability is small. Monte Carlo simulations of walkers alternating between two Lévy-like step-length distributions support the experimental findings. Our results extend the study of Lévy flights to a class of nonideal random walks, with length-dependent local exponents and alternating step-length distributions, that are likely to be relevant in other real physical systems.
\end{abstract}
\maketitle

\section{Introduction}

Lévy flight is a kind of random walk with a step-length distribution with divergent variance. Lévy flights are often modeled by a step-length distribution that decays asymptotically as a power law $P(\ell)\sim \ell^{-1-\alpha}$, with the Lévy index $\alpha\leq 2$. The main feature of Lévy flight is the occurrence of large steps that, although rare, dominate the dynamics of the random walk. For example, the large steps lead the transmission by a sample of thickness $L$ to decay as $L^{-\alpha/2}$, slower than that for normal diffusion \cite{Barthelemy2008,Baudouin2014,Buldyrev20012,Klinger2022}. Lévy flights are present in a variety of systems, just to name a few: animal foraging \cite{Viswanathan1996}, phonons transport in nanowires \cite{Li2022} and spread of genes lineage \cite{Smith2023}.

Multiple scattering of light by atomic resonant vapor can be described as a Lévy flight random walk \cite{Pereira2004,Zaburdaev2015,Mercadier2009}, where the step length $\ell$ is the distance traveled by a photon between successive scatterings. In each scattering event, a photon is absorbed and subsequently re-emitted by an atom. Due to the thermal motion of the atoms (Doppler effect) and to elastic collisions that may perturb the atomic dipole during the scattering process, the re-emitted photon generally has a frequency different from the absorbed one: the light's frequency is thus redistributed at each scattering event. Since the interaction cross-section decreases rapidly with the detuning from the atomic resonance, a photon re-emitted in the wings of the spectral line travels a long distance before being scattered again. These rare off-resonance emissions are responsible for the long steps typical of Lévy flights. Experimental investigations of Lévy flight of light in atomic vapor have been concentrated on determining the Lévy index $\alpha_T$ from transmission measurements \cite{Baudouin2014,Araujo2021,Macedo2021,Lopez2023}. The step-length distribution for light in atomic vapor is not a power law \cite{Pereira2007,Chevrollier2012,Lopez2023,Nunes2025} but could be modeled as:
\begin{equation}
	P(\ell)\sim \ell^{-1-\alpha(\ell)},
	\label{Eq:Levy_Power} 
\end{equation}
with the Lévy index $\alpha(\ell)$ slowly varying with the step length. The smooth variation of $\alpha(\ell)$ with $\ell$ allows us to describe the step-length distribution locally as a power law with local exponent $\alpha(\ell)$. Measurements of $\alpha_T$ from transmission have evidenced a dependence on the size of the system \cite{Macedo2021,Lopez2023}. Moreover, the steps alternate between two different step-length distributions, related to two different cases of frequency redistribution \cite{Hummer1962,Domke1988,Pereira2007}, depending on whether or not a collision between atoms has occurred during the photon scattering process.

In the present article, we investigate experimentally the transmittance of light by a resonant atomic cesium vapor. Our aim is twofold: i) to determine if the local exponent of the step-length distribution is retrieved from transmission measurements and ii) to obtain information of how the alternation between two distinct step-length distributions influences the transmission through a sample. Simulations are made to confirm the experimental results. 

The article is organized as follows: in section II, we discuss the frequency redistribution, giving emphasis to the existence of two distinct cases, whether there are or not collisions between atoms during the scattering event. Furthermore, we describe how the frequency redistribution influences the step-length distribution of the random walk. In section III, we describe the experiment and discuss the experimental results. In section IV, we describe simulations of Lévy flight random walk with length-dependent power law exponent and two distinct distributions. In section V, we compare and discuss the results of experiment and simulations and, finally, conclude in section VI.

Throughout this article we use two distinct symbols for the Lévy index, which differ only in the way the index is obtained. The step-length-dependent Lévy index $\alpha(\ell)$, defined by Eq. \ref{Eq:Levy_Power}, is obtained directly from the step-length distribution and depends on the step length. The Lévy index obtained by fitting the measured or simulated diffuse transmission to a power law, as in done experimentally in section III, is denoted $\alpha_T$. For an ideal Lévy flight, whose step-length distribution has a single index, the two coincide, $\alpha_T=\alpha$ \cite{Buldyrev20012,Klinger2022}. When the local exponent varies with the step length they need not, and determining at which step length the fitted index $\alpha_T$ samples $\alpha(\ell)$ is one of the aims of this work.

\section{Frequency redistribution for light scattered by an atomic vapor}

The light scattered by resonant alkali atomic vapor has its frequency redistributed by two main mechanisms: (i) Doppler shifts and (ii) atomic elastic collisions that randomize the atomic dipole phase \cite{Hindmarsh1973,Allard1982,Pereira2007}. This frequency redistribution results in a broadened scattered spectrum with each frequency interacting with the vapor with a different cross-section. The larger the detuning to atomic resonance, the lower is the cross-section, and the higher is the probability of large distance traveled  before a new scattering event. The step-length distribution is related to the interaction cross-section $\sigma(x')$ and to the scattered spectral profile $\Theta(x')$ by:
\begin{equation}
P_n(\ell)=\beta_0\int dx'\,\Theta_n(x')\,\frac{\beta(x')}{\beta_0}\,e^{-N\sigma(x')\ell},
	\label{Eq.P(l)}
\end{equation}
where $x'$ denote the scattered detuning normalized by Doppler width, $N$ is the atomic density and $\beta(x')=N\sigma(x')$ is the absorption coefficient, and $\beta_0=\beta(x'=0)$. Note that $\ell$ denotes the modulus of the distance travelled between two successive scattering events.

It is worth emphasizing the physical content of Eq.~\ref{Eq.P(l)}. For a photon of fixed detuning $x'$, the free path between two successive scattering events follows the Beer-Lambert law: the step-length distribution conditioned on the frequency is the exponential $P(\ell|x') = \beta(x')e^{-\beta(x')\ell}$, with mean free path $1/\beta(x')$. However, due to the frequency redistribution, the photon is re-emitted at each scattering event with a new detuning $x'$ drawn from the scattered spectral profile $\Theta_n(x')$. The unconditioned step-length distribution $P_n(\ell)$ is therefore a continuous superposition of Beer-Lambert exponentials weighted by $\Theta_n(x')$, as expressed in Eq.~\ref{Eq.P(l)}. This shows that $\ell_0=\beta^{-1}_0$, the mean step length for photons scattered at the line center, is the typical length scale of the transport. It is the shortest such scale, since the absorption coefficient is maximum at resonance. Since the absorption coefficient $\beta(x')$ varies by orders of magnitude between the line center and the wings of the spectral line, this superposition of exponentials with very different decay rates results in a heavy-tailed, non-exponential distribution, which can be locally approximated by the power law of Eq.~\ref{Eq:Levy_Power}. The underlying hypotheses are thus that the vapor is homogeneous, so that each individual step at fixed frequency is governed by the Beer-Lambert exponential attenuation, and that the photon frequency is redistributed at each scattering event.

The scattered spectral profile depends on the scattering event number, $n$, and on the incident detuning $x$ and can be calculated using a recurrence rule \cite{Mercadier2013,Nunes2025}:
\begin{equation}
	\Theta_{n}(x')=\int dx C(x)\Theta_{n-1}(x)\frac{R(x',x)}{\phi(x)}
	\label{Eq.:Recurrence},
\end{equation}
with $\Theta_{n-1}(x)$ the scattered profile at step $n-1$, $C(x)$ a system size dependent extinction factor that computes the probability of the photon with detunig $x$ to be scattered by the vapor. $\phi(x)$ is a normalized absorption profile ($\int dx \phi(x)=1$) and $R(x,x')$ is a joint frequency redistribution function \cite{Hummer1962}. Hummer \cite{Hummer1962}, has calculated the joint redistribution function for four cases, two of which will be considered here for scattering by alkali vapor: (i) case II, for pure natural broadening (no collisions), and (ii) case III, for combined natural and collisional broadening. Both cases II and III are also Doppler broadened in the laboratory rest frame. We use the redistribution function $R(x,x')$ proposed by Hummer \cite{Hummer1962} that are averaged over all angles between incident and scattered photons in three dimensional space.

For finite vapor system, $\Theta_n(x')$ stabilizes after some scattering events \cite{Nunes2025} with the scattering profile for case $II$ converging to a Doppler profile and case $III$ converging to a Voigt profile with heavy tailed Lorentzian wings, that favors the occurrence of large steps \footnote{The comment on the convergence to Doppler profile or Voigt profile for cases II and III, respectively, is a simplification of the scattering dynamics to keep the main features of each case. A more detailed discussion involving the effect of the size of the system and the atomic electronic structure can be found in \cite{Nunes2025}}. The frequency redistribution function can be written as a weighted average \cite{Omont1972,Domke1988,Post1986,Vermeersch1988}:
\begin{equation}
	R(x,x')=(1-P_C)R_{II}(x,x')+P_CR_{III}(x,x'),
\end{equation}
with $P_C=\frac{\Gamma_C}{\Gamma_C+\Gamma_n}$ being the probability that a collision between atoms occurs during a scattering event \cite{Post1986,Vermeersch1988}. $\Gamma_C$ and $\Gamma_n$ denote collisional and natural widths, respectively.

The photons can undergo some scattering events before a collision between the atoms occurs. During its random walk, the photons intercalate between the two distinct step-length distributions corresponding to $P_{II}(\ell)$ for scattering with no collision between atoms (case $II$) and $P_{III}(\ell)$ if a collision occurs during the scattering (case $III$).
\begin{equation}
	P(\ell)=\begin{cases}
	P_{II}(\ell),  & 
	\begin{array}{@{}l@{}}
	\text{if no collisions occurs} \\
	\text{during scattering}
	\end{array} \\
	P_{III}(\ell), & 
	\begin{array}{@{}l@{}}
	\text{if collision occurs} \\
	\text{during scattering.}
	\end{array}
	\end{cases}
\end{equation}

\begin{figure}
	\centering
	\includegraphics[scale=1]{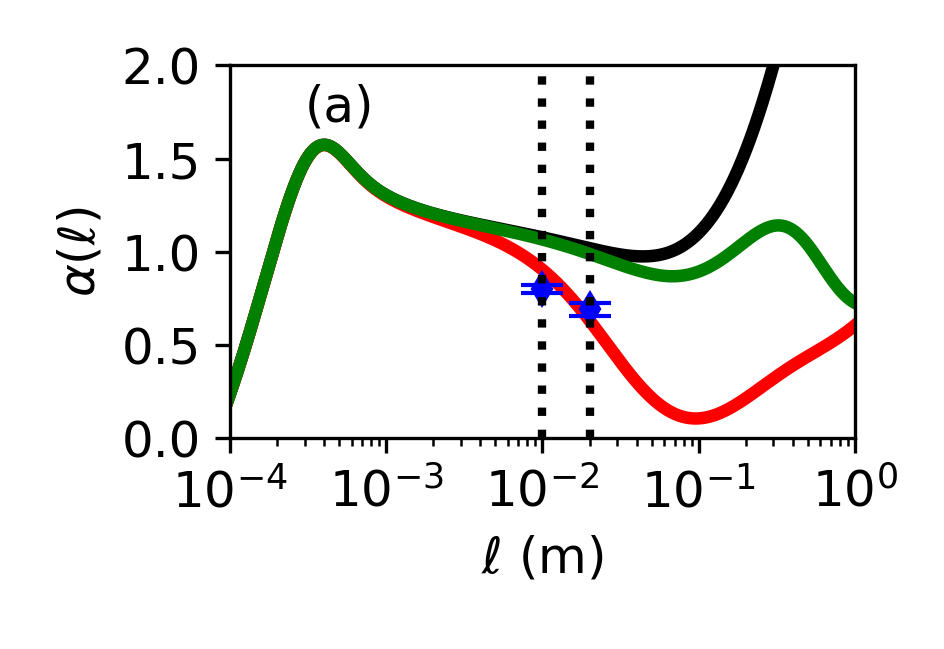}\\
	\vspace{-.6cm}
	\includegraphics[scale=1]{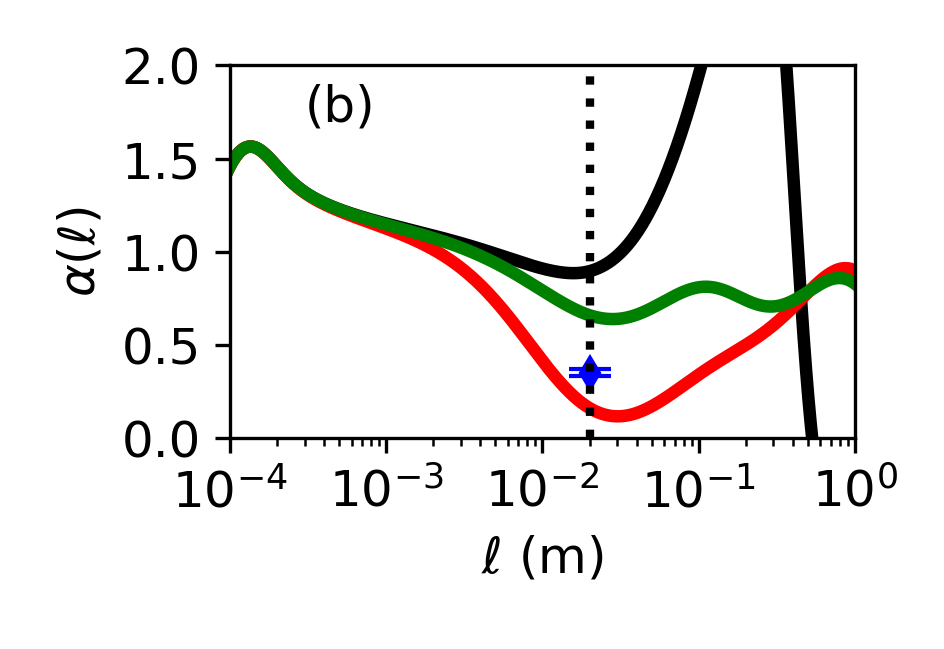}\\
	\vspace{-.6cm}
	\includegraphics[scale=1]{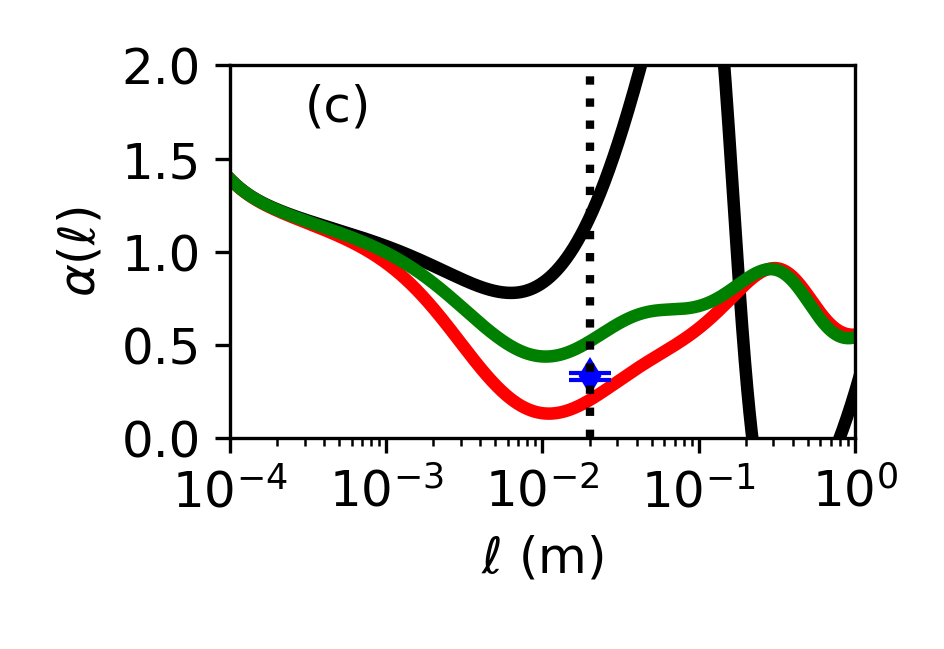}\\
	\vspace{-.6cm}
	\includegraphics[scale=1]{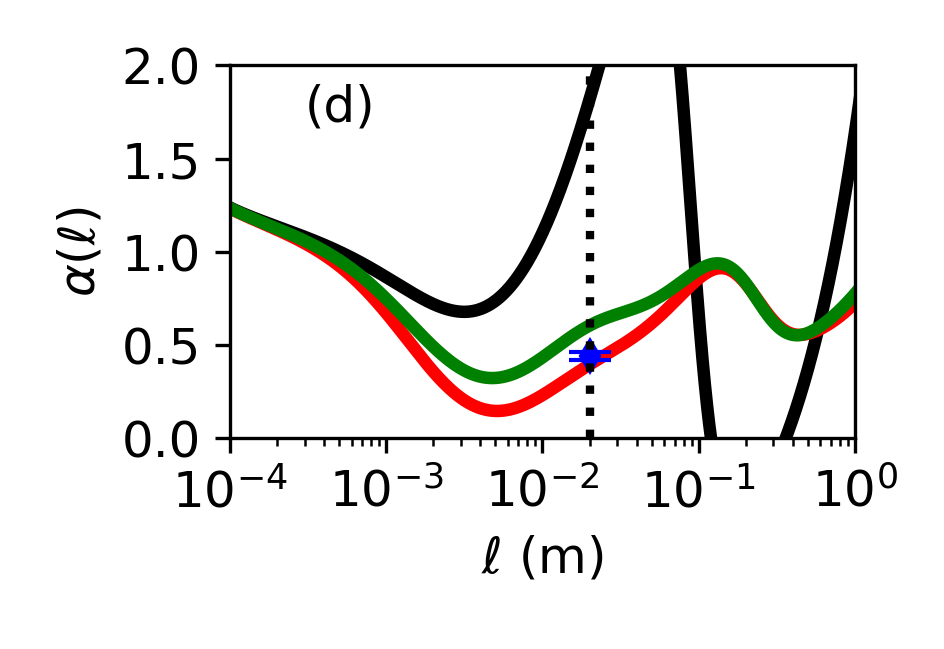}
	\vspace{-.6cm}
	\caption{Measured Lévy index $\alpha_T$ for different atomic densities (diamonds). Lines are step-length-dependent Lévy index $\alpha(\ell)$ calculated from $P_{II}(\ell)$ (black lines), $P_{III}(\ell)$ (red lines) and $P(\ell)$ (green lines).  (a) For density $N=3.5\times 10^{12}$ atoms/cm$^3$ corresponding to $P_C=6\%$; (b) for $N=10\times 10^{12}$ atoms/cm$^3$ and $P_C=15\%$;  (c) for $N=25\times 10^{12}$ atoms/cm$^3$ and $P_C=30\%$;  (d) for $N=44\times 10^{12}$ atoms/cm$^3$ and $P_C=43\%$. For density (a) measurement were made for two different cells of thickness $L=1$ cm and $L=2$ cm. For densities (b-d) only the cell of thickness $L=2$ cm was used.}
	\label{fig:Resultado_Exp_1}
\end{figure}

From the calculated step-length distributions and from Eq. \ref{Eq:Levy_Power}, we can extract the local power law exponent $1+\alpha(\ell)=-\frac{dlog_{10}(P(\ell))}{dlog_{10}(\ell)}$. In Fig. \ref{fig:Resultado_Exp_1} we show $\alpha(\ell)$ calculated from $P_{II}(\ell)$, $P_{III}(\ell)$ and $P(\ell)$ for different atomic densities. Looking, for instance, at Fig. \ref{fig:Resultado_Exp_1}(a) we see that $\alpha(\ell)$ for case II varies smoothly with $\ell$ keeping in a range $1.0\leq \alpha\leq 1.2$ for a wide range of step lengths (from 2 mm to 3 cm). For case $III$, $\alpha$ decreases to values as low as $\alpha\sim 0.1$ related to the transition from the Doppler core of the scattering profile to the Lorentzian wings and increases to $\alpha\sim 0.5$ for larger steps related to the Lorentzian wings of the scattering profile \cite{Pereira2004,Chevrollier2012,Lopez2023,Nunes2025}. The $\alpha(\ell)$ calculated from the total step-length distribution $P(\ell)$ in Fig. \ref{fig:Resultado_Exp_1}(a) ($P_C\sim 6\%$) approaches the case II for steps with $\ell<10^{-1}$ m, as collisions between atoms are rare. However, for large steps, the $P(\ell)$ distribution approaches the $P_{III}(\ell)$ distribution, since there is a cut-off in the $P_{II}(\ell)$ distribution \cite{Pereira2007,Nunes2025} consequence of the scattering coherence in the atomic rest frame.

The Lévy index $\alpha(\ell)$ is a property of the step-length distribution: it depends on the step length and on the parameters of the vapor --- atomic density, Voigt parameter and collision probability $P_C$. It retains only a residual dependence on the size of the system, through the extinction factor $C(x)$ of Eq. \ref{Eq.:Recurrence}, which accounts for the escape of far-wing photons from a finite sample.

\section{Experiment}

\begin{figure}
    \centering
    \includegraphics[width=.9\linewidth]{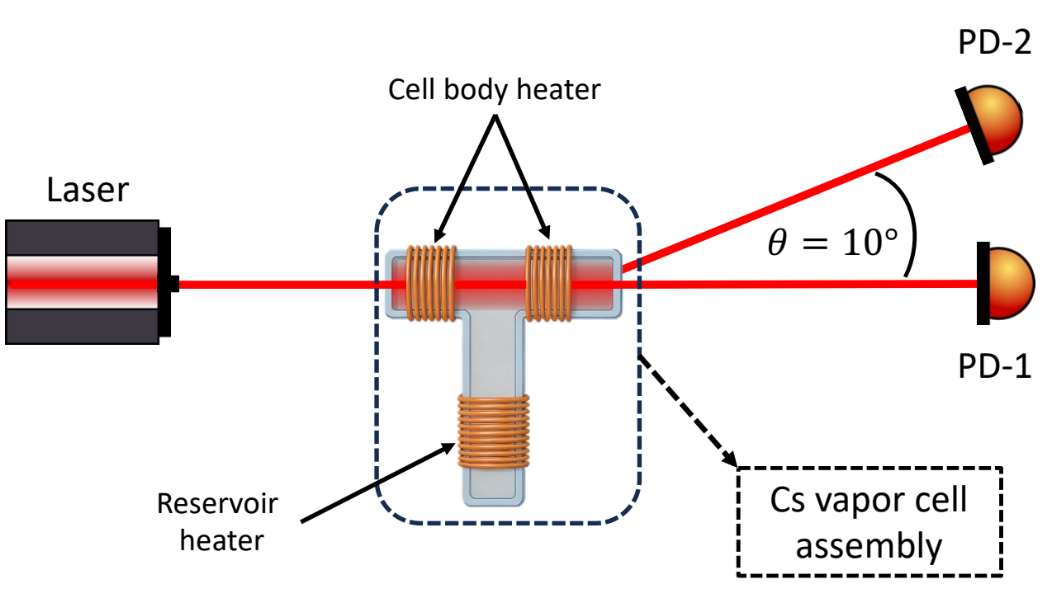}
    \caption{Scheme of the experimental set-up. A low power laser beam is incident on the atomic vapor and diffuse transmission if collected by photodetector PD-2 at 10$^\circ$ relative to laser beam. Photodetector PD-1 is used to detect ballistic transmission for determining atomic density.}
    \label{fig:setup}
\end{figure}
The experimental set-up is shown in Fig \ref{fig:setup}. A laser beam with wavelength of 852 nm is incident on a cesium vapor, contained in a sealed cell, and scanned around the D$_2$ line. Two independent ovens heat the cell's probed volume and the liquid cesium reservoir. The density of the vapor is increased by heating the reservoir such that the optical density at the line center is a minimum of 60; that is, there is no ballistic transmission around the line center. The light might undergo multiple scattering events before escaping the vapor and is collected as a diffuse transmission by photodetector PD-2. Photodetector PD-2 is at an angle of 10$^\circ$ related to the laser beam axis and is placed past a lens to collect diffuse light. More details of the experimental parameters are given in the appendix. 

Our experimental aim is to measure the diffuse transmission for fixed atomic density $N$ and cell thickness $L$ as a function of the starting point of the random walk $z_0$. Diffuse transmission is expected to scale as \cite{Buldyrev20012,Klinger2022}:
\begin{equation}
	T_D\propto \left(\frac{z_0}{L}\right)^{\alpha_T/2}. 
	\label{Eq.Klinger}
\end{equation}

Equation \ref{Eq.Klinger} is the asymptotic scaling of an ideal Lévy flight of index $\alpha$, for which $\alpha_T=\alpha$. Since the step-length distribution of light in a vapor has no single index, we do not assume this scaling to hold globally: Eq. \ref{Eq.Klinger} is used as a local fitting relation over a restricted range of $z_0$, and $\alpha_T$ is by construction a local logarithmic slope.

\begin{figure}
	\centering
	\includegraphics[scale=1.]{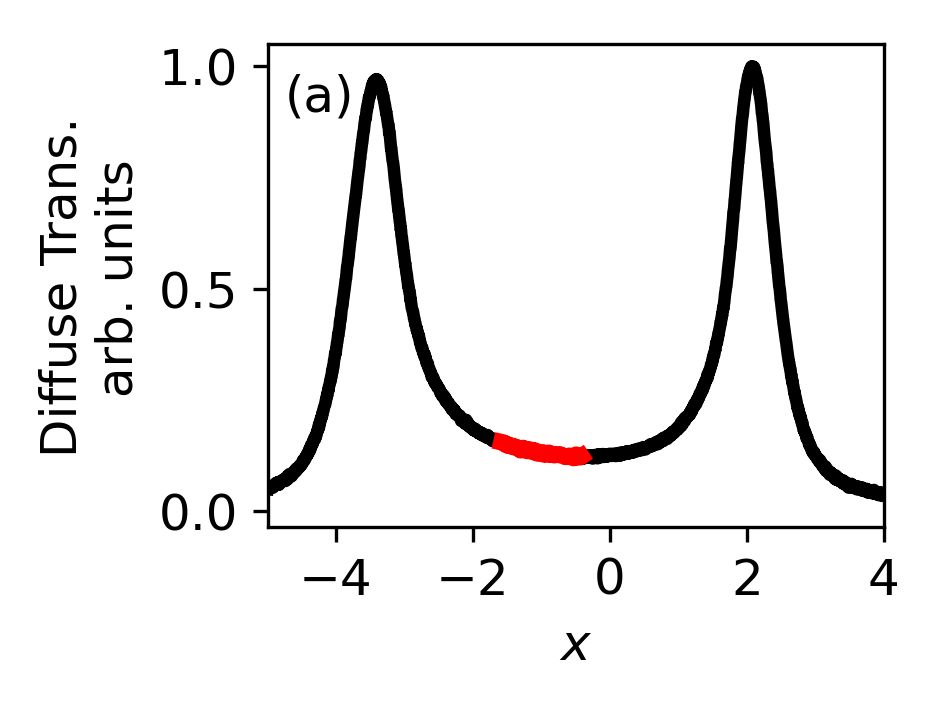}\\
	\includegraphics[scale=1.]{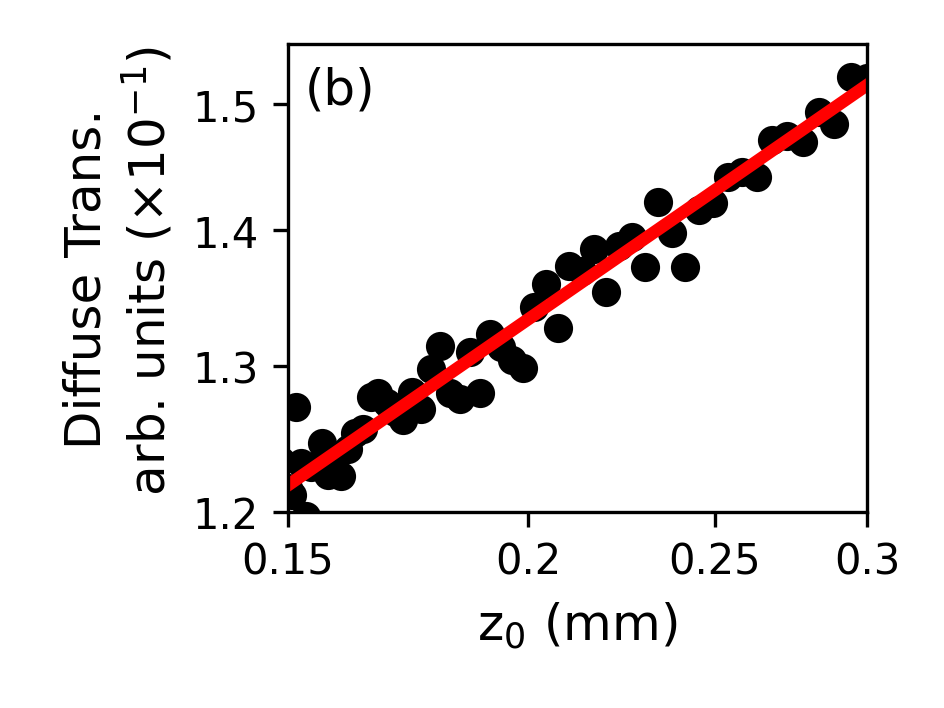}\\
	\caption{(a) Diffuse transmission spectrum for a cell of thickness $L=2$ cm and atomic density of $N=3.5\times10^{12}$ atoms/cm$^3$. The zero detuning corresponds to the $6S_{1/2}(F=4)\rightarrow 6P_{3/2}(F'=5)$ transition. (b) Power law dependence of diffuse transmission as a function of laser penetration depth $z_0$. The range of frequencies analysed in (b) corresponds to red-solid line in (a).}
    \label{fig:Spectrum}
\end{figure}

In our experiment, the laser frequency is scanned around the $6S_{1/2}(F=4)\rightarrow 6P_{3/2}(F'=3,4,5)$ cesium transitions with the starting point of the random walk defined as the penetration depth of the laser beam given by $z_0=\left(N\sigma(x)\right)^{-1}$ \cite{Lopez2023}. Note that $z_0$ is the mean free path of the incident photon, evaluated at the incident laser frequency, and it defines the starting point of the random walk through the penetration depth of this photon --- whose well-defined frequency corresponds to a Dirac delta incident spectral profile in Eq.~\ref{Eq.P(l)}, so that this penetration follows the simple exponential (Beer--Lambert) law. Only after the first scattering event is the frequency redistributed according to $\Theta_1(x')$, and the subsequent steps are governed by the heavy-tailed distribution of Eq.~\ref{Eq.P(l)}.

An example of diffuse transmission spectrum is shown in Fig. \ref{fig:Spectrum}. The frequency axis of the spectrum in Fig \ref{fig:Spectrum}(a) can be transformed into the penetration depth axis to build Fig. \ref{fig:Spectrum}(b), which shows a power law dependence of diffuse transmission as a function of the starting point of the random walk as predicted in Eq. \ref{Eq.Klinger}. Diffuse transmission is fitted by a power law to obtain the Lévy index $\alpha_T$. The results of measured $\alpha_T$ for two different cells of thickness $L=1$ cm and $L=2$ cm for the same atomic density of $N=3.5\times10^{12}$ atoms/cm$^3$ are shown as diamonds in Fig. \ref{fig:Resultado_Exp_1}(a). The results for three other densities for $L=2$ cm are shown in Figs. \ref{fig:Resultado_Exp_1}(b-d)\footnote{Theoretical curves depend on the absorption Voigt profile parameter $a=(\Gamma_C+\Gamma_n)/\Gamma_D$ and thus depend on the atomic density, such that different theoretical curves should be drawn for each density.}

Two main results are evidenced in Fig. \ref{fig:Resultado_Exp_1}: i) The measured values $\alpha_T$ depend on the system size and follow the curve $\alpha(\ell)$; ii) the measured $\alpha_T$ follow the expected curve $\alpha(\ell)$ for the case $III$, that is, for scattering with collisions between atoms even if the collisional width is small relative to the natural width ($\Gamma_C\sim 0.1\Gamma_n$ for Fig. \ref{fig:Resultado_Exp_1}(a)).

\section{Simulations}

\subsection{Double Pareto distribution}

{In order to isolate the two features of interest --- a step-length distribution with a varying local power exponent, and the alternation between two distributions --- we first consider a deliberately minimal model, in which a random walk is simulated in one dimension following the distributions $P^{(DP)}(\ell)$:
\begin{align}
        P^{(DP)}(\ell)=\begin{cases} P^{(P)}_{II}(\ell), & \text{with probability $1-P_C$}\\
        P^{(P)}_{III}(\ell), & \text{with probability $P_C$}\end{cases},
\end{align}
with $P^{(P)}_{(II,III)}(\ell)$ being normalized Pareto distributions:
\begin{equation}
    P^{(P)}_{(II,III)}(\ell)=\frac{\alpha_{(II,III)}}{\ell_0}\left(\frac{\ell_0}{\ell}\right)^{1+\alpha_{(II,III)}}{H}(\ell-\ell_0), 
    \label{Eq.Pareto}
\end{equation}
with H the Heaviside function and $\ell_0$ the lower cutoff of the power law. The simulations are performed with dimensionless quantities: all lengths ($\ell,z_0,L$) are measured in units of $\ell_0$, so that we set $\ell_0=1$ without loss of generality. The Lévy indexes of the distributions are chosen such that $1<\alpha_{II}<2$ and $0<\alpha_{III}<1$ to mimic the ranges of the local exponent $\alpha(\ell)$ calculated for the two step-length distributions of light diffusing in atomic vapor, corresponding to Doppler and Voigt scattering profiles (cases II and III, see Fig. 1), respectively. In the simulations reported below we have used $\alpha_{II}=1.4$, $\alpha_{III}=0.3$ and $P_C=1\%$. The system sizes $L$ and starting points $z_0$ are chosen so as to probe the two asymptotic regimes discussed in the literature: the \textit{continuous limit} $\ell_0\ll z_0\ll L$ \cite{Klinger2022} and the opposite limit $z_0\ll\ell_0\ll L$ \cite{Klinger2022}.

\begin{figure}
	\centering
	\includegraphics[scale=1.]{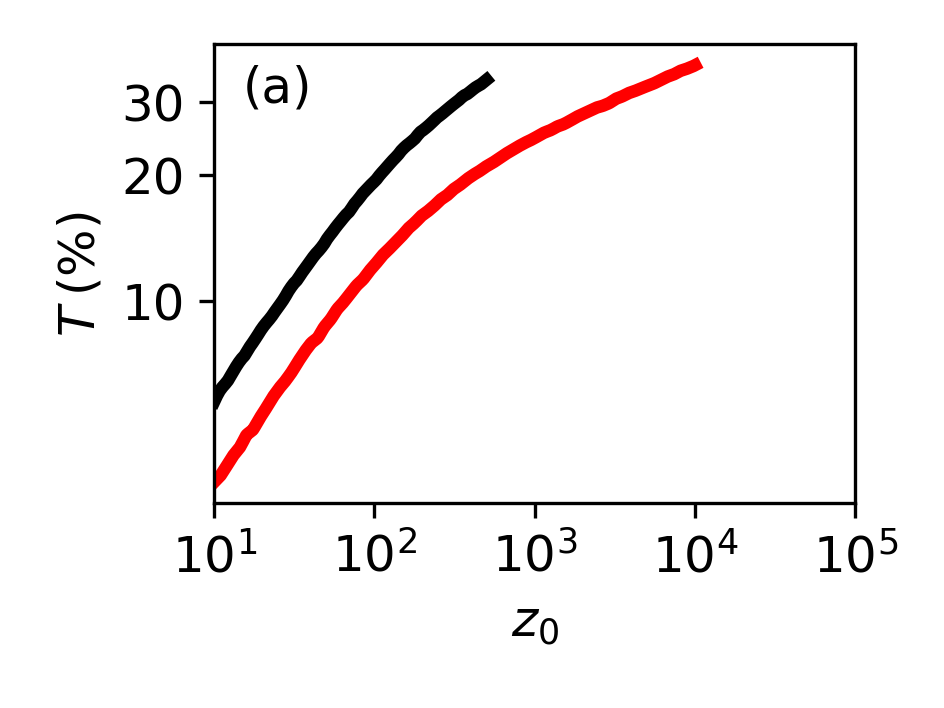}\\
	\includegraphics[scale=1.]{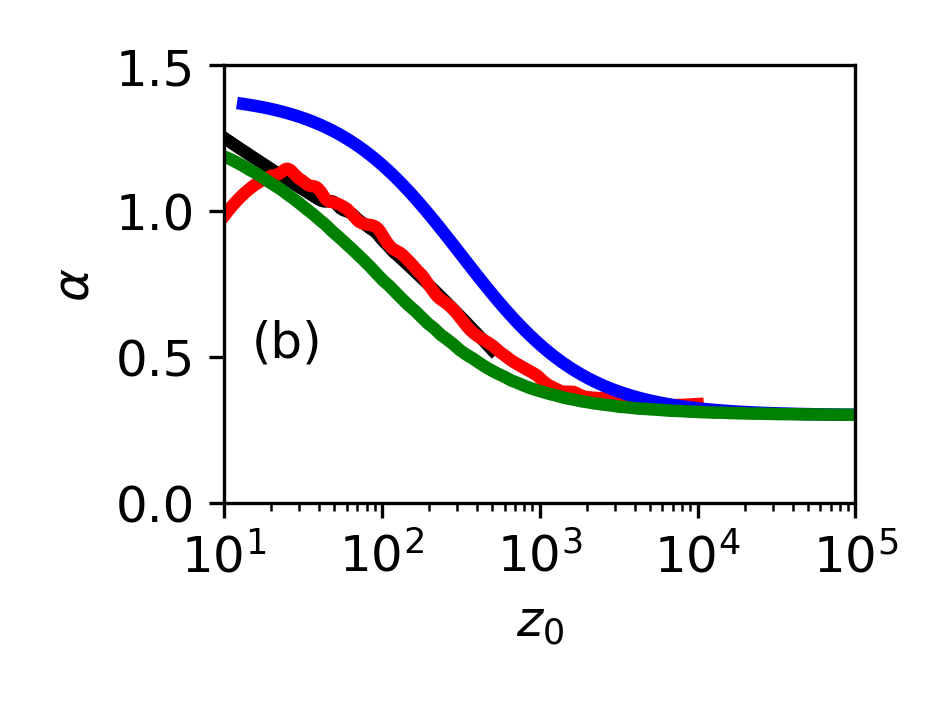}
	\caption{(a) Transmission as a function of starting point $z_0$ (in units of $\ell_0=1$) for two different sizes of system: $L=5\times 10^{3}$ (black-solid line) and $L=10^5$ (red-solid line). (b) Lévy index $\alpha_T$ obtained by fitting locally Eq. \ref{Eq.Klinger} for $L=5\times 10^{3}$ (black-solid line) and $L=10^5$ (red-solid line). In blue-solid line we plot expected $\alpha$ value calculated from the step-length distribution: $1+\alpha=-\frac{d log(P^{(DP)})}{dlog(z_0)}$. In green-solid line we plot expected $\alpha^{(w)}$ obtained from Eq. \ref{Eq:alfa_Weigthed}.}
    \label{fig:Duplo_Pareto_z0}
\end{figure}

\subsubsection{Transmission as a function of starting point}

The walker evolves in a 1D space $0<z<L$ with open boundaries at $z=0$ and $z=L$ in the sense that a walker crossing either boundary leaves the interval and is not allowed to return \cite{Buldyrev20012,Klinger2022}, with starting point of the random walk $z_0$ obeying $\ell_0\ll z_0\ll L$, the so-called \textit{continuous limit} \cite{Klinger2022}. We run the simulation for fixed $\ell_0$ and $L$ for various starting points $z_0$. We collect the transmission, $T$, through the boundary at $z=L$ as a function of $z_0$, see Fig. \ref{fig:Duplo_Pareto_z0}(a). Then, using Eq. \ref{Eq.Klinger}, we determine the local slope of $log_{10}(T)$ as a function of $log_{10}(z_0)$ to obtain $\alpha_T/2$. The obtained values of $\alpha_T$ using this procedure for two different system sizes are shown in Fig. \ref{fig:Duplo_Pareto_z0}(b). We have run the simulation with $\ell_0=1$, $L=5\times 10^{3}$ and $L=10^5$ and $z_0$ ranging from 10 to $L/10$. 

The obtained values of $\alpha_T$ depend on the starting point $z_0$ and do not depend on the size of the system $L$. As the distance from the starting point to the boundary at $z=0$ is much less than the distance to $z=L$ the majority of the walkers escape the system on the left side. As $T+T^{(left)}=1$ the transmission by $z=L$ is strongly influenced by the number of particles that escape by $z=0$ (denoted by $T^{(left)}$), which should depend on $z_0$ and very little on the large system size $L$. We conclude that $T$ is governed by $T^{(left)}$, which depends on the starting point $z_0$ and not on the size of the system $L$. Hence, the obtained values of $\alpha_T$ from the simulations should be related to the local slope of $P^{(DP)}$ around $z_0$, calculated from $1+\alpha=-\frac{d log(P^{(DP)})}{dlog(z_0)}$, which we plot as a blue line in Fig. \ref{fig:Duplo_Pareto_z0}. The local slope follows quite well the dependence of $\alpha_T$ obtained from the simulations, which reaffirms the argument that $T^{(left)}$ rules the dynamics. 

From the simulations, we can compute the number of walkers that have escaped by $z=0$ after a step corresponding to $P^{(P)}_{II}$ or $P^{(P)}_{III}$. We denote those numbers by $N_{II}$ and $N_{III}$, respectively. We have estimated a weighted $\alpha^{(w)}_T$ value by:
\begin{equation}
    \begin{split}
    \alpha^{(w)}_T(z_0)=\frac{N_{II}(z_0)}{N_{II}(z_0)+N_{III}(z_0)}\alpha_{II}
    \\
    +\frac{N_{III}(z_0)}{N_{II}(z_0)+N_{III}(z_0)}\alpha_{III}.
    \end{split}
    \label{Eq:alfa_Weigthed}
\end{equation}
$\alpha^{(w)}_T$ is plotted as a green-solid line in Fig. \ref{fig:Duplo_Pareto_z0}. It follows quite well the obtained $\alpha_T$ values. 

\subsubsection{Transmission as a function of the microscopic length scale $\ell_0$}

We have also tested in the simulations the regime for which $z_0\ll \ell_0\ll L$ \cite{Klinger2022} by fixing $z_0$ to $10^{-3}$ and running simulations for $\ell_0$ ranging from $1$ to $5$ \footnote{Note that for each value of $\ell_0$ there is a different step-length distribution defined in Eq. \ref{Eq.Pareto}. For this reason, we change $\ell_0$ over a small range to keep the step-length distribution approximately the same. }. The simulated diffuse transmission $T$ as a function of $\ell_0$ is shown in Fig. \ref{fig:TD_r0}(a) for different system sizes. It is clear that the diffuse transmission follows the power law given by \cite{Klinger2022}:
\begin{equation}
    T\propto \left(\frac{\ell_0}{L}\right)^{\alpha_T/2}. 
    \label{Eq.TD_r0}
\end{equation}
From Eq. \ref{Eq.TD_r0} we can extract $\alpha_T$ from the simulated $T$. The $\alpha_T$ values clearly depend on the size of the system, and we plot the $\alpha_T$ values obtained from simulations in Fig. \ref{fig:TD_r0}(b) together with the local exponent of the step-length distribution for step of size $\ell=L$, calculated from $1+\alpha(\ell=L)=-\left.\frac{dlog_{10}(P(\ell))}{dlog_{10}(\ell)}\right|_{\ell=L}$. The obtained $\alpha_T$ from the simulations closely follow the local exponent of the step-length distribution for $\ell=L$. Also plotted in Fig. \ref{fig:TD_r0}(b) is a weighted $\alpha^{(w)}_T$ value taken as:

\begin{equation}
    \alpha^{(w)}_T=\frac{N_{II}(\ell_0)}{N_{II}(\ell_0)+N_{III}(\ell_0)}\alpha_{II}+\frac{N_{III}(\ell_0)}{N_{II}(\ell_0)+N_{III}(\ell_0)}\alpha_{III}, 
    \label{Eq:Weigthed2}
\end{equation}
with $N_{II}$ and $N_{III}$ the number of walkers transmitted by $z=L$ after performing a step corresponding to $P^{(P)}_{II}$ and $P^{(P)}_{III}$ distributions, respectively.

\begin{figure}
    \centering
    \includegraphics[scale=1.]{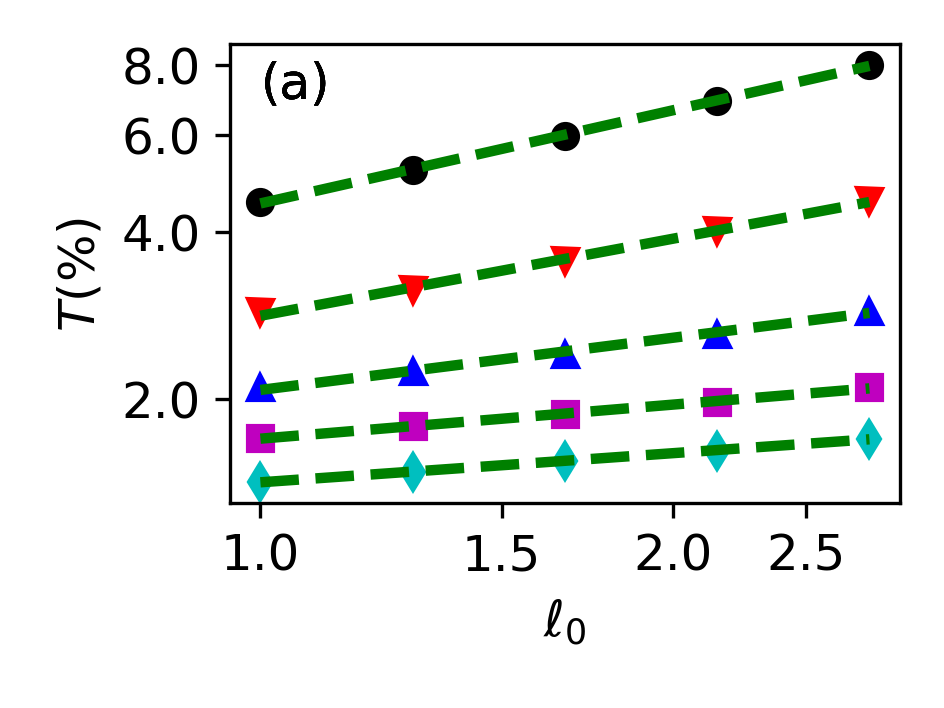}\\
    \includegraphics[scale=1.05]{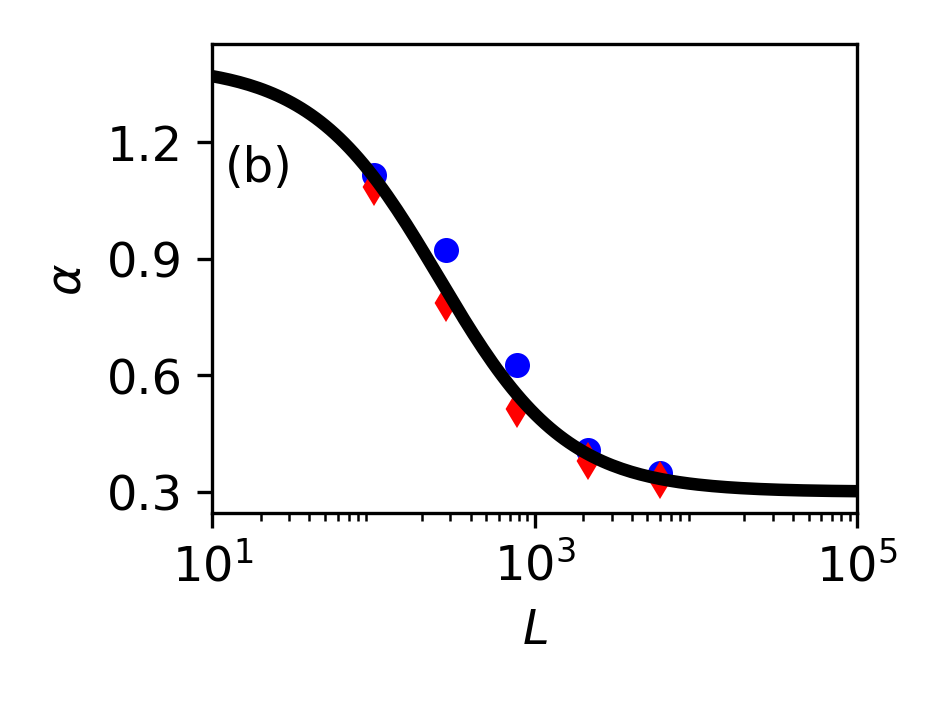}
    \caption{(a) Diffuse transmission as a function of $\ell_0$ for different sizes of systems $L$: $L=100$ (black circles), $L=280$ (red triangle), $L=780$ (blue triangle), $L=2150$ (magenta squares) and $L=6000$ (cyan diamonds). (b) Extracted Lévy index $\alpha_T$ by fitting Eq. \ref{Eq.TD_r0} to transmission as a function of $L$ (blue circles). Black-solid line: expected $\alpha$ values calculated from $1+\alpha(\ell)=-\frac{dlog_{10}(P(\ell))}{dlog_{10}(\ell)}$. Red-diamonds: Expected $\alpha^{(w)}$ calculated using Eq. \ref{Eq:Weigthed2}.}
    \label{fig:TD_r0}
\end{figure}

\subsection{Photon diffusion in atomic vapor}

\subsubsection{Description of the simulation}

We simulate the photon diffusion in an atomic vapor using the Monte Carlo method. For simplification of the simulation routine, we consider a two-level atom and work with dimensionless lengths, expressed in units of a length scale $\ell_0$ set by the atomic density. We consider a slab geometry with longitudinal axis $z$ and open boundaries at $z=0$ and $z=L$: the vapor simply ends at those planes, so that a photon crossing either of them leaves the medium and never returns. The simulation follows the sequence of steps described below. The photon is incident along the $z$ axis and is first scattered at position $z_0$. We denote by $\vec{v}$ the velocity of the scattering atom normalized by the most probable speed $u=\sqrt{2k_BT/M}$, and by $\hat{k}$ the unit vector along the current photon propagation direction. As absorption is velocity selective, the dimensionless atomic velocity component parallel to the photon direction, $v_{\parallel}=\vec{v}\cdot\hat{k}$, is not distributed according to a Maxwell–Boltzmann law for a photon at detuning $x$ in the laboratory frame, but is instead drawn from \cite{Anderson1995,Carvalho2021}:
\begin{equation}
    P(v_\parallel)\propto \frac{a^2}{a^2+(x-v_\parallel)^2}e^{-v^2_\parallel},
\end{equation}
where $a$ is the Voigt parameter defined as the ratio of homogeneous and Doppler width. The velocity components perpendicular to the incident photon direction are taken from a two-dimensional Maxwell-Boltzmann distribution. The photon detuning in the atomic rest frame, $x^{at}$, Doppler shifted from the laboratory frame detuning $x$, is calculated as:
\begin{equation}
    x^{at}=x-v_\parallel.
    \label{dessintonia_ref_atomico}
\end{equation}

The emission in the atomic rest frame depends on the occurrence of a collision during the scattering with probability $P_C$. If no collision occurs, the scattered and incident detuning are the same in the atomic rest frame $x'^{at}=x^{at}$. If a collision occurs, the scattered detuning in the atomic rest frame is taken from a Lorentzian distribution \cite{Hummer1962,Domke1988}:
\begin{equation}
	P(x'^{at})= \frac{a}{\pi}\frac{1}{(x'^{at})^2+a^2}.
	\label{distribuicao_lorentziana}
\end{equation}

Then the scattered direction $\hat{n}'$ is isotropically drawn over $4\pi$ sr and the scattered detuning is Doppler shifted back to the laboratory rest frame:
\begin{equation}
    x'=x'^{at}+\hat{n}'\cdot \vec{v}=x'^{at}+v_\parallel\cos\theta+v_\perp\sin\theta
    \label{decomposicao_velocidades}
\end{equation}
where $\theta$ is the angle between the incident and the scattered photon directions and $v_\perp$ is the component of $\vec{v}$ along the direction perpendicular to $\hat{k}$ lying in the plane defined by $\hat{k}$ and $\hat{n}'$ \cite{Carvalho2021}. Note that Eqs. \ref{dessintonia_ref_atomico} and \ref{decomposicao_velocidades} combine into $x'=x+\vec{v}\cdot(\hat{n}'-\hat{k})$ whenever the scattering is coherent in the atomic rest frame (case II), which is the standard three-dimensional expression for the Doppler contribution to frequency redistribution. The two frame transformations are kept separate because for case III the rest-frame detuning is redrawn from Eq. \ref{distribuicao_lorentziana} and the compact expression no longer holds. Once we have the scattered detuning we draw the distance traveled by the photon until the next scattering event following the distribution:
\begin{equation}
    P_S(\ell)=\beta(x')e^{-\beta(x')\ell}. 
    \label{Eq.BeerLambert}
\end{equation}

Sampling Eq. \ref{Eq.BeerLambert} with the scattered detuning $x'$ drawn as described above reproduces the step-length distribution of Eq. \ref{Eq.P(l)}: the exponential law holds at fixed frequency, and the average over the scattered spectral profile is performed by the Monte Carlo procedure itself. The position of the photon is then updated. If $z>L$, the photon has left the sample through the far boundary and is computed as transmitted; if $z<0$, it has left through the entrance boundary and is discarded; if $0<z<L$, the photon is still inside the vapor and the process is restarted. Because of the long-tailed step-length distribution, a photon typically crosses a boundary plane in a single large step rather than reaching it, so that what is imposed is a first-crossing condition rather than a condition on the photon density at the boundary \cite{Buldyrev20012,Klinger2022}. The absorption coefficient is a Voigt profile defined as \cite{Pereira2007}:
\begin{equation}
    \beta(x)=\frac{1}{\ell_0}\frac{a}{\pi^{3/2}}\int_{-\infty}^{\infty}\frac{e^{-y^2}}{a^2+(x-y)^2}dy.
\end{equation}
The length $\ell_0$ is thus fixed by the atomic density, since $\beta\propto N\sigma$.

\subsubsection{Transmission as a function of starting point $z_0$}

We simulate the photon random walk and collect diffuse transmission as a function of incident detuning, as in the experiment. The incident photon, at detuning $x$, penetrates the vapor a distance $z_0$ given by the distribution of Eq. \ref{Eq.BeerLambert}. In Fig. \ref{fig:Simulacao_Variand_F} we plot diffuse transmission as a function of the mean penetration depth $\bar{z}_0=1/\beta(x)$ on a log-log scale. As expected, a power law behavior is obtained \cite{Buldyrev2001,Klinger2022,Lopez2023} (see Eq. \ref{Eq.Klinger}) from which we extract the Lévy index $\alpha_T$. However, the obtained values of $\alpha_T$ are in the range $0.9<\alpha_T<1.0$ and do not depend on the size of the system $L$ even for the change of $L$ by a factor of 10. The fact that the power dependence of $T$ with $z_0$ does not depend on size of the system is similar to that found for the simulation with the Double Pareto distribution. 

\begin{figure}
    \centering
    \includegraphics[scale =1.]{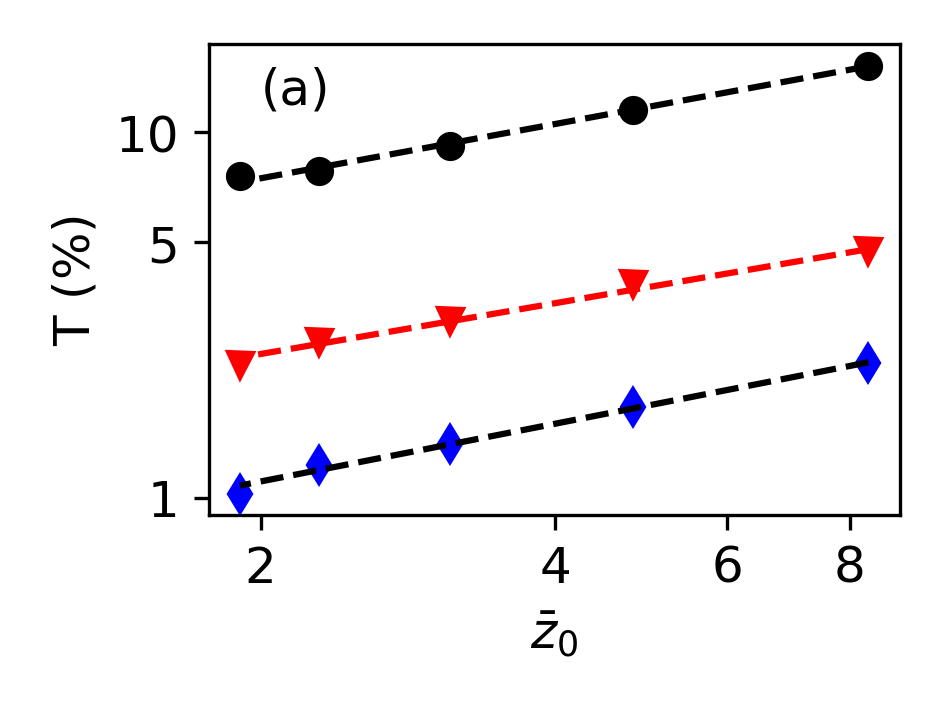}\\
    \includegraphics[scale=1.]{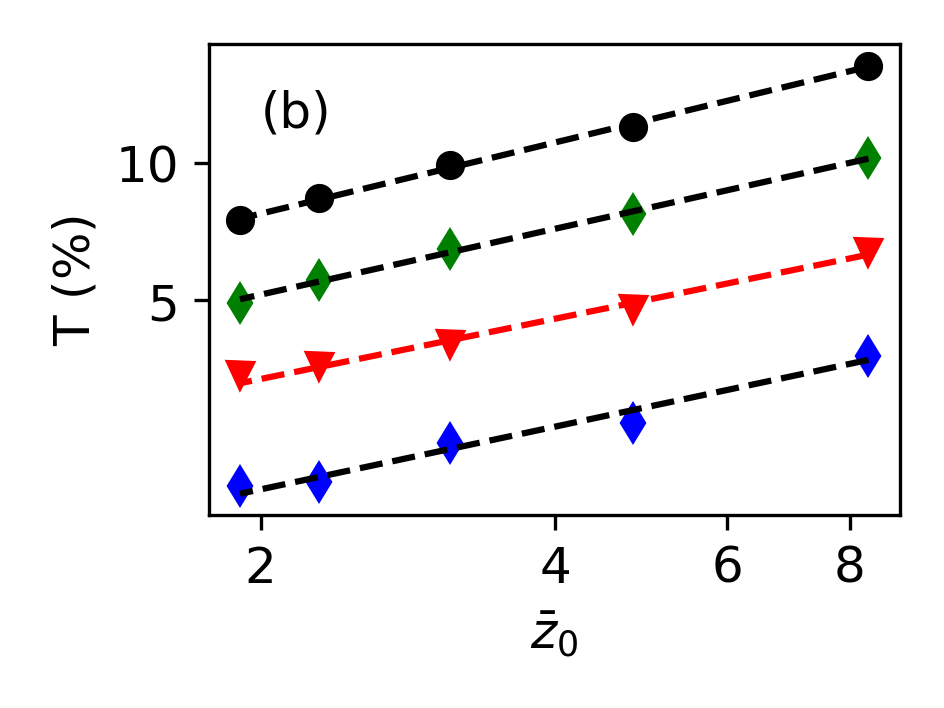}
    \caption{Diffuse transmission as a function of mean penetration depth, $\bar{z}_0$ (in units of $\ell_0=1$), for four different cell thicknesses: $L=120$ (black-circles), $L=270$ (green diamonds), $L=540$ (red triangles) and $L=1080$ (blue diamonds). The diffuse transmission are fitted by Eq. \ref{Eq.Klinger} to obtain the Lévy index $\alpha_T$. (a) for $P_C=0$ and (b) for $P_C=100\%$.}
    \label{fig:Simulacao_Variand_F}
\end{figure}

\subsubsection{Transmission as a function of microscopic length scale}

We also simulate the diffuse transmission as a function of the ratio $\ell_0/L$. Since only this ratio is relevant, we vary it by changing the system size $L$ at fixed $\ell_0=1$, rather than by changing $\ell_0$. In an experiment, the same ratio is instead changed by adjusting the atomic density, as was done in \cite{Baudouin2014,Araujo2021,Macedo2021}. We perform the simulation in the limit $z_0\ll \ell_0$, for which the diffuse transmission scales as Eq. \ref{Eq.TD_r0}. We extract the Lévy index $\alpha_T$ by fitting diffuse transmission as a power law (Eq.~\ref{Eq.TD_r0}). The extracted Lévy index is then compared in Fig.~\ref{fig:Sim_fotons_lo} with theoretical predictions for $\alpha(\ell)$ values corresponding to the local exponent of the step-length distribution (Eq.~\ref{Eq.P(l)}). Note that the system sizes used here differ from those of Fig.~\ref{fig:Simulacao_Variand_F}, as the two sets of simulations are independent.

\begin{figure}[h!]
    \centering
\includegraphics[scale=.95]{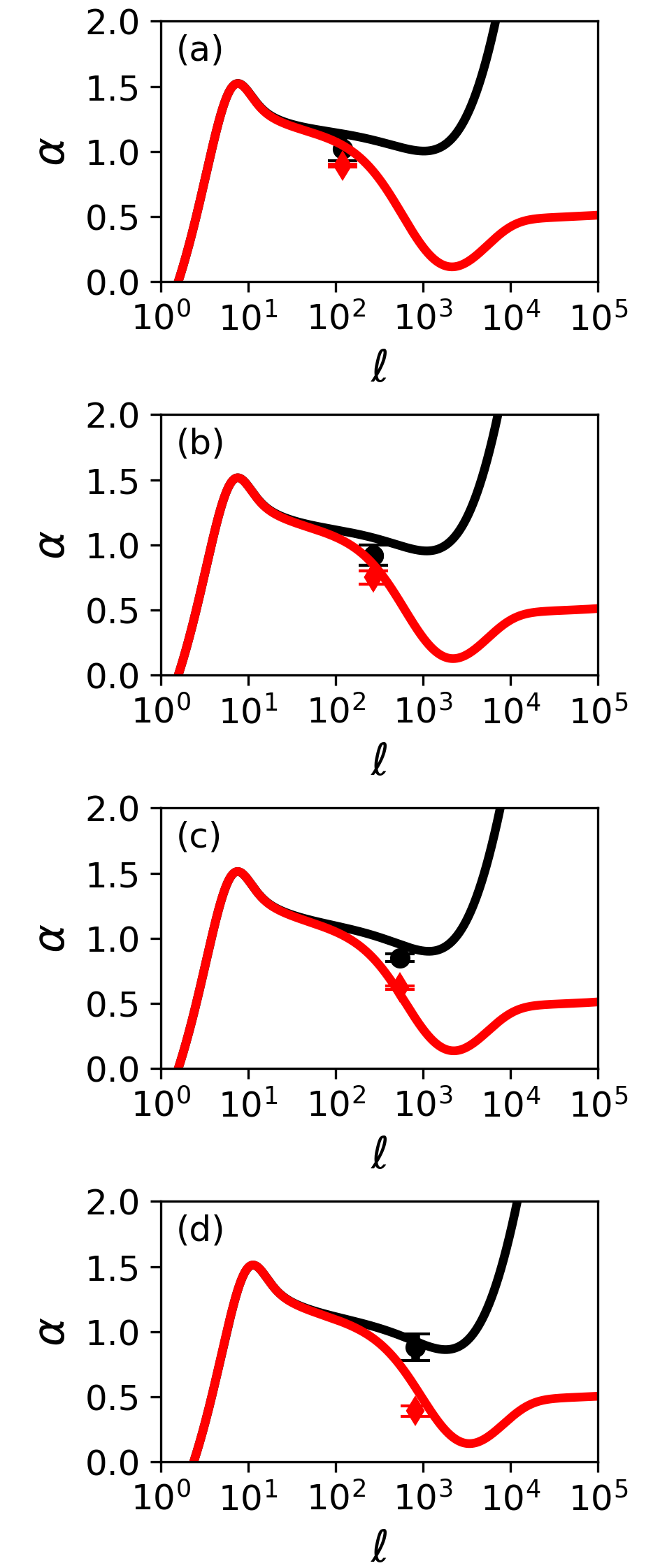}\\
	\caption{Calculated local power exponent of step-length distributions for $P_C=0$ (black-solid line) and $P_C=100\%$ (red-solid line) for $\ell_0=1$ and different cell thicknesses (a)$L=120$, (b) $L=270$, (c)$L=540$ and (d) $L=810$. In each Fig., symbols are $\alpha_T$ values extracted from fitting Eq. \ref{Eq.TD_r0} to simulated diffuse transmission. In each panel the symbols are plotted at $\ell=L$, so that the agreement between symbols and curves is a direct verification of $\alpha_T=\alpha(\ell=L)$.} 
	\label{fig:Sim_fotons_lo}
\end{figure}

The extracted $\alpha_T$ values from the simulations follow well the theoretically predicted curves of $\alpha(\ell)$. We can clearly distinguish between the two distinct step-length distributions corresponding to coherent scattering in the atomic rest frame (case $II$) and incoherent scattering (case $III$). Moreover, the $\alpha_T$ values depend on the system size $L$ with the value extracted from the simulations corresponding well to $\alpha(\ell=L)$, a result similar to that obtained in the experiment.
\begin{figure}
    \centering
    \includegraphics[scale=1.]{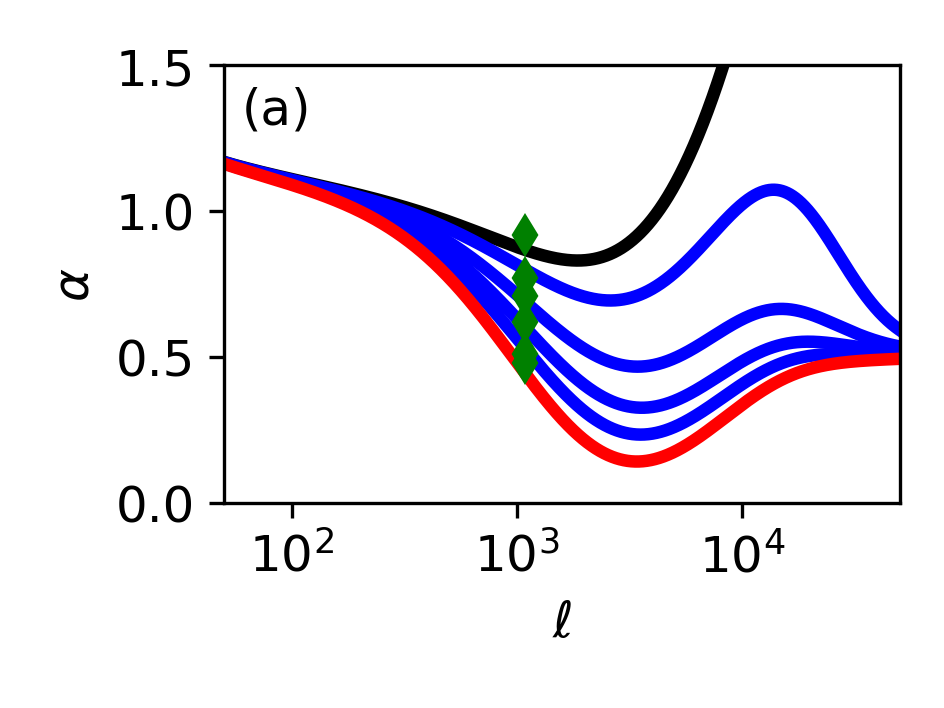}\\
    \includegraphics[scale=1.]{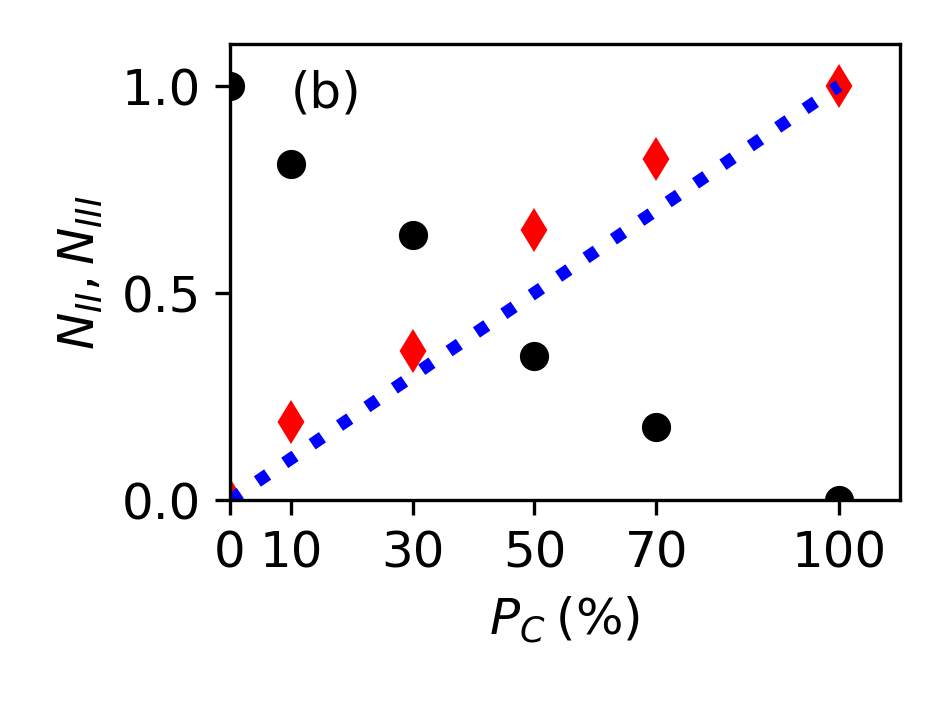}\\
    \includegraphics[scale=1.]{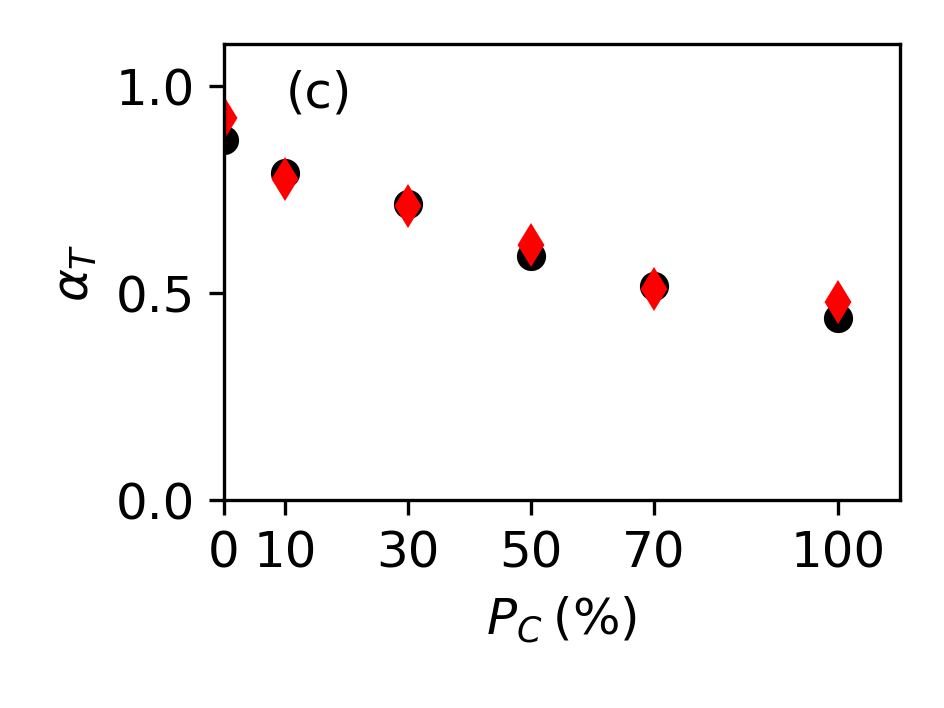}\\
    \caption{(a) Calculated power-law local exponents as a function of step-length for different collision probabilities $P_C$ (0\%, 10\%, 30\%, 50\%, 70\% and 100\%, from top to bottom). Symbols are $\alpha_T$ values obtained from Monte Carlo simulations. (b) Percentage of photons that have escape after step in case II ($N_{II}$, black circles) and case (III) ($N_{III}$, red diamonds) as a function of probablity of collisions, $P_C$. Blue-dotted line corresponds to $N_{II,III}=P_C$. (c) Comparison of $\alpha_T$ values obtained from simulated transmissions (red diamonds) compared to estimated values of $\alpha^{(w)}_T$ (black circles), weighted with $N_{II}$ and $N_{III}$.}
    \label{fig:Simulacao_PC}
\end{figure}

To see the effect of different probability of collisions, we have obtained $\alpha_T$ as a function of $P_C$ for a system of size $L=1100$ from simulations. In Fig. \ref{fig:Simulacao_PC}(a) we show theoretically calculated $\alpha(\ell)$ curves for $P_C=$ 0\%, 10\%, 30\%, 50\%, 70\% and 100\% as well as the $\alpha_T$ values from simulation transmissions. The simulated values follow well the theoretical predictions. We also obtain from the simulations the percentage of photons that escape by $z=L$ after a step of case II (III), denoted as $N_{II}$($N_{III}$), in a way similar to section IV-A, which are plotted in Fig. \ref{fig:Simulacao_PC}(b). The percentage of photons that escape after a step corresponding to case $III$ is higher than $P_C$ as the large steps are more common in case $III$. We estimated a weighted Lévy index as $\alpha^{(w)}_T=N_{II}\alpha_{II}+N_{III}\alpha_{III}$, with $\alpha_{II}$($\alpha_{III}$) theoretical local exponent for case II(III) for $\ell=L$, which follows very well the extracted $\alpha_T$ from the simulations (see Fig. \ref{fig:Simulacao_PC}(c)).

\section{Discussion}

\subsection{On the simulations}
\begin{figure}
    \centering
    \includegraphics[scale=1.]{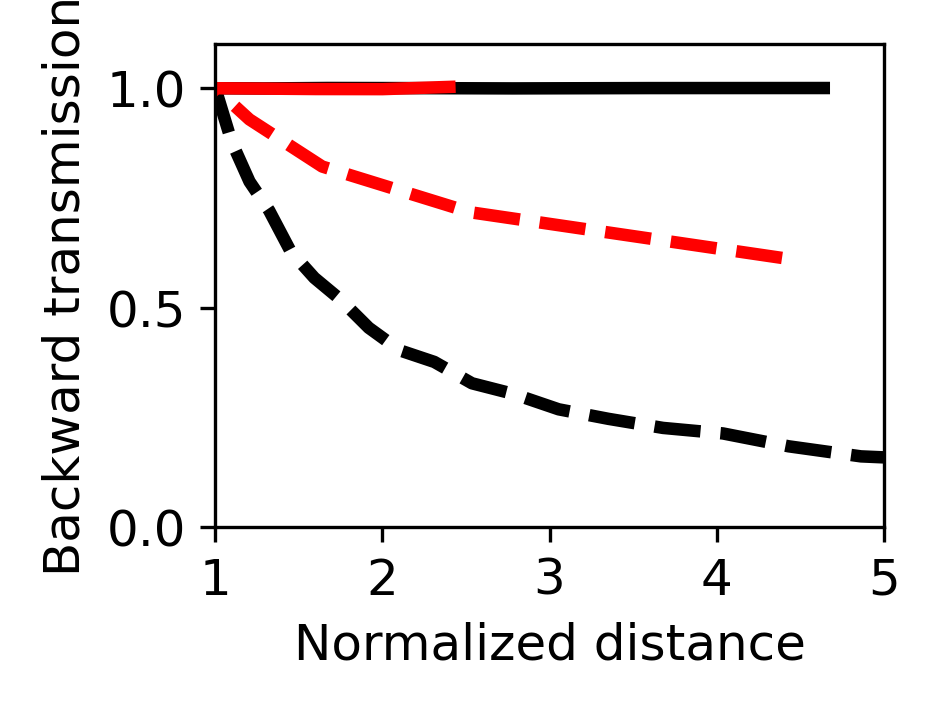}
    \caption{Simulated backward transmission after the first scattering event. Black lines are for Double Pareto step-length distributions and red lines are for photon diffusion in atomic vapor. Solid lines are for diffuse transmission as a function of $\ell_0$ and dashed lines are for diffuse transmission as a function of $z_0$. Distance are normalized by the smaller value of $\ell_0$ and $z_0$ and transmission is normalized by the value at minimum $\ell_0$ or $z_0$.} 
    \label{fig:Backward}
\end{figure}
Two different results were found depending on the simulations: i) for transmission as a function of $z_0$ in the limit $\ell_0\ll z_0$, the extracted $\alpha_T$ values do not depend on the size of the system $L$; ii) for transmission as a function of $\ell_0$ in the limit $z_0\ll \ell_0$, the extracted $\alpha_T$ values depend on the size of the system and follow $\alpha_T=\alpha(\ell=L)$. The two cases have distinct behavior relative to the backward escape for the first step as we show in Fig. \ref{fig:Backward} for both Double Pareto and photon diffusion simulations. In case (i), the backward escape for the first step is very sensitive to the variable used $z_0$ and the forward transmission is ruled by this backward escape. As $z_0\ll L$, it is natural to think that the forward transmission does not depend on $L$. In contrast, in case (ii), as $\ell_0$ is much higher than the starting point $z_0$, the dependence of the backward escape after first step with the variable $\ell_0$ is very smooth and $\alpha_T$ depends on the system size $L$.

\subsection{Comparison between simulation and experiment}

In the experiment, we have varied the starting point of the random walk $z_0$, through the penetration depth of the laser, as a function of incident detuning. As detuning $x\neq 0$, the experiment corresponds to case (i) of the discussion in the previous section for which simulations have not shown a dependence of $\alpha_T$ with the size of the system. However, experiment has shown dependence with both cell length ($L=1$ cm and $L=2$ cm) and effective system size ($L=2$ cm for different atomic densities, see also \cite{Lopez2023}). Further investigations are needed to better understand those results. In simulations, $\alpha_T$ depends on the system size $L$ if backward escape after first scattering is, almost, independent of the variable. Eventually, a mechanism not yet understood in the experiment might turn backward escape independent of incident detuning. Ideally, experimental investigation would require, a larger variety of cell thicknesses for the same density (more experimental points in Fig. \ref{fig:Resultado_Exp_1}). However, we disposed of only two good cells to perform the experiment.

\subsection{System-size dependence of the transmission power law exponent}

The central result of both experiment and simulation can now be stated without ambiguity: $\alpha_T=\alpha(\ell=L)$, that is, the Lévy index obtained from the transmission is the local index of the step-length distribution evaluated at the scale set by the size of the system. The step-length distribution itself carries no such dependence; the system size enters through the length at which its local index is sampled. It is worth stressing that the three lengths appearing in this discussion have distinct status. The system size $L$ is fixed by the cell; the starting point $z_0$ is the mean penetration depth of the incident beam and is scanned experimentally through the laser detuning; and $\ell_0$ is the microscopic length scale of the medium, set by the atomic density. None of them is a mean free path of the walk: for a Lévy flight with $\alpha\leq1$ the mean step length diverges, so that no such scale exists \cite{Mercadier2009, Araujo2021, Macedo2021}.

\subsection{On the experimental value of $P_C$}

One curious result of the experiment is that the obtained values of $\alpha_T$ correspond to case III even though $P_C\ll 1$ (for instance, $P_C=6\%$ for Fig. \ref{fig:Resultado_Exp_1}(a)). The value of $P_C$ was estimated by calculating the collisional width as proportional to the density $\Gamma_C=kN$ with $k=9\times10^{-8}\:\mathrm{Hz}\cdot \mathrm{cm}^{3}$ \cite{Jabbour1995,Meng2016,Silans2018}. The density was estimated from fitting a theoretical curve to ballistic transmission (detected with photodetector PD-1, see Fig. 1). However, impurities might be present in a vapor cell increasing the value of $P_C$. We have performed saturated absorption experiments to check that our cells have not a large amount of impurities and measured a homogeneous width of $\Gamma=6.8\pm 0.3$ MHz ($\Gamma=1.3\Gamma_n$) at room temperature. Taking into account the extra width of $0.3\Gamma_n$ for the calculations of $\Gamma_C$ the values of $P_C$ would be of 27$\%$ for the density used in Fig. \ref{fig:Resultado_Exp_1}(a) and of 52$\%$ for the density used in Fig. \ref{fig:Resultado_Exp_1}(d). Those new values of $P_C$ are not high enough to justify that the measured $\alpha_T$ is close to the case $III$ curve, and it is still quite surprising from the atomic physics point of view to observe such a large effect of collisions ($\alpha_T$ values consistent with case III) even for $\Gamma_C<\Gamma_n$.

\section{Conclusions}

We have investigated experimentally the light diffusion by a resonant atomic vapor which can be treated as a Lévy flight random walk \cite{Pereira2004,Mercadier2009,Mercadier2013,Chevrollier2012}. Emphasis is given to the fact that the walker alternates between two distinct step-length distributions, namely $P_{II}(\ell)$ and $P_{III}(\ell)$, which corresponds to case $II$ and $III$ of frequency redistribution, in the light scattering by vapor described by Hummer \cite{Hummer1962}. Each of the step-length distribution can be described locally as a power law with step-length-dependent Lévy index $\alpha(\ell)$. We have obtained the Lévy index $\alpha_T$ for different system sizes $L$ and atomic densities from transmission measurement. The measured values correspond to those expected for a step with the length of the system size: $\alpha_T=\alpha(\ell=L)$. 

We have simulated random walk with step-length distribution with local exponent that depends on the length of the step: (i) using a Double Pareto distribution and (ii) simulating the photon diffusion in resonant vapor. Our emphasis in the simulations is to observe if the transmission power law exponent $\alpha_T$ depends on the size of the system as in the experiment. Three parameters were relevant in the discussions: a typical step length $\ell_0$, the starting point of the random walk $z_0$ and the size of the system $L$. Analyzing the diffuse transmission as a function of $z_0$ for the limit $\ell_0\ll z_0\ll L$ \cite{Buldyrev20012,Klinger2022} (continuous limit) no dependence of $\alpha_T$ with system size was found. Analysing the diffuse transmission as a function of $\ell_0$ for the limit $z_0\ll \ell_0\ll L$ \cite{Klinger2022}, we extracted from simulation a system-size-dependent Lévy index, as in the experiment, that is, $\alpha_T=\alpha(\ell=L)$. The simulation results were similar for both step-length distributions proposed: Double Pareto and photon diffusion in atomic vapor.

The random walk for light in atomic vapors can be described as alternating between two distinct step-length distributions $P_{II}(\ell)$ and $P_{III}(\ell)$ with the first occurring with probability $(1-P_C)$ and the last with probability $P_C$ ($P_C$ being the probability of atomic collisions during a light scattering event). The simulation of photon diffusion on atomic vapor has shown that $\alpha_T$ depends on $P_C$. The extracted $\alpha_T$ value depends on the relative number of walkers that have escape the system after a step in $P_{II}(\ell)$ or $P_{III}(\ell)$.

In the experiment, the measured values of $\alpha_T$ are consistent with those expected for case $III$ even for $P_C\ll 1$. For the atomic physics community it is intriguing to observe a signal related to atomic collisions (case $III$) even for $\Gamma_c\ll\Gamma_n$. Also, the experiment was analyzed by collecting diffuse transmission as a function of $z_0$ and a system-size-dependent $\alpha_T$ was measured. Further investigations are needed to understand the differences between experiment and simulations results, with probably a factor influencing the experiment not taken into consideration. In particular, the emergence of Case III characteristics in a regime of negligible collision probability ($P_C \ll 1$) challenges the standard interpretation of frequency redistribution and suggests an underlying mechanism that has yet to be fully understood. 

This article presents a real physical system described by Lévy flight random walk that is more complicated than those usually modeled in the literature: local power exponent dependence on the step length and alternation between two Lévy-like step-length distributions. We are not aware of investigations of Lévy flights for systems with similar characteristics in the literature. As a perspective for future simulation work, the sharp lower cutoff of the Pareto distribution could be avoided by employing smoother heavy-tailed distributions, such as the generalized Pareto distribution \cite{Tommasi2024} recently used in Monte Carlo studies of anomalous radiative transfer, which contains the exponential (Beer-Lambert) case as a limit and for which the proper treatment of boundary crossings in finite media has been established. We hope that both the experimental and simulation results presented here will encourage further investigation on similar non-usual Lévy flights random walk systems that might be important in non-ideal real systems.

\begin{acknowledgments}
T.P.d.S. and J.P.L. acknowledge financial support from Conselho Nacional de Desenvolvimento Científico e Tecnológico, Brazil (CNPq, Public Call CNPq/MCTI/FNDCT  N$^\circ$ 18/2021), Financiadora de Estudos e Projetos, Brazil (FINEP, N$^\circ$ 01.13.0235.07). I.C.N. acknowledge financial support from Coordenação de Aperfeiçoamento de Pessoal de Nível Superior, Brazil (CAPES).
\end{acknowledgments}

\section*{DATA AVAILABILITY}
The data that support the findings of this article are not publicly available. The data are available from the authors upon reasonable request.

\section*{Appendix: details of the experiment}

In the experiment, we used two distinct cells: one with thickness of $L=1$ cm and the other with $L=2$ cm. Both are glass cells with cylindrical shapes. The $L=1$ cm cell has a diameter of $4.5$ cm while the $L=2$ cm cell has a diameter of $2$ cm.

Each cell is heated by two independent ovens, one that determines the temperature of the vapor and the other that determines the temperature of a side-arm reservoir containing liquid cesium. The side-arm temperature determines the vapor pressure, while the vapor temperature determines the Doppler width and the Voigt parameter.

From experimental conditions of Fig. \ref{fig:Resultado_Exp_1}(a) to Fig. \ref{fig:Resultado_Exp_1}(d), side-arm temperature was rised to increase atomic density with a slight increase of vapor temperature with the Doppler width ranging from $\Gamma_D=250$ MHz for density of $N=3.5\times10^{12}$ atoms/cm$^3$ (Fig. \ref{fig:Resultado_Exp_1}(a)) to $\Gamma_D=268$ MHz for density of $N=44\times 10^{12}$ atoms/cm$^3$ (Fig. \ref{fig:Resultado_Exp_1}(d)) corresponding to Voigt parameters values of $a=10^{-2}$ and $a=2\cdot 10^{-2}$, respectively. 

The laser used in the experiment is a semiconductor laser with external grating and is scanned around the $6S_{1/2}(F=4)\rightarrow 6P_{3/2}(F'=3,4,5)$ transitions. The beam has a diameter of 1.3 mm and its power is within the range of 1 to 9 $\mu$ W corresponding to intensities of 0.1 to 0.9 times the saturation intensity. For the intensity range used, the laser-vapor interaction is linear and no change in $\alpha_T$ with intensity was observed. The analyzes of diffuse transmission as a function of incident detuning are made in the range $-1.6 \leq x\leq -0.4$. The analyses are performed at the red side of the $6S_{1/2}(F=4)\rightarrow 6P_{3/2}(F'=5)$ transition to avoid the contribution of photons detected after a single scattering event (see \cite{Lopez2023} for more details). The limit $x\leq -0.4$ is to ensure a monotonic growth of $T_D$ as a function of $z_0$ by avoiding the maximum of absorption coefficient around $x=-0.3$. The limit $-1.6\leq x$ ensures a effective frequency redistribution around the line center  \cite{Post1986,Vermeersch1988,Lopez2023}.

\bibliographystyle{apsrev4-2}
\bibliography{LevyBib.bib}

\end{document}